\documentstyle[12pt]{article}
\parskip=\medskipamount

\newcommand {\cA}{{\cal A}}

\newcommand {\cD}{{\cal D}}

\newcommand {\cF}{{\cal F}}
\newcommand {\cG}{{\cal G}}
\newcommand {\cH}{{\cal H}}

\newcommand {\cL}{{\cal L}}

\newcommand {\cU}{{\cal U}}

\newcommand {\cW}{{\cal W}}

\newcommand{\bB}{{\bf B}}

\def\a{\alpha}
\def\b{\beta}

\def\d{\delta}

\def\f{\phi}

\def\G{\Gamma}

\def\l{\lambda}

\def\o{\omega}

\def\q{\theta}
\def\r{\rho}
\def\s{\sigma}
\def\t{\tau}

\def\z{\zeta}

\def\F{\Phi}
\def\J{\Psi}
\def\L{\Lambda}

\newcommand{\ad}{{\dot{\alpha}}}                           %new
\newcommand{\bd}{{\dot{\beta}}}                            %new
\newcommand{\pa}{\partial}                           %new
\newcommand{\hf}{\frac12}

\newcommand{\be}{\begin{equation}}
\newcommand{\ee}{\end{equation}}
\newcommand{\bea}{\begin{eqnarray}}
\newcommand{\eea}{\end{eqnarray}}
\newcommand{\non}{\nonumber}
\begin{document}
%%%%%%%%%%%%%%%%%%%%%%%%%
%%%%%%%%%%%%%%%%%%%%%%%%

\begin{titlepage}
\thispagestyle{empty}

\begin{flushright}
TP-TSPU-2/98\\
TP-TSU-7/98 \\
hep-th/9804168 \\
\end{flushright}

%\vspace{1cm}
\begin{center}
{\Large\bf  Comments on the Background Field Method \\
in Harmonic Superspace:}\\
{\large\bf Non-holomorphic Corrections in N = 4 SYM
}
\end{center}
%\vspace{3mm}

\begin{center}
{\bf Ioseph L. Buchbinder}\\
\footnotesize{
{\it
Department of Theoretical Physics,
Tomsk State Pedagogical University\\
Tomsk 634041, Russia}\\
\tt{josephb@tspi.tomsk.su}
}
%\vspace{3mm}
\end{center}

\begin{center}
{\bf Sergei M. Kuzenko}\\
\footnotesize{
{\it
Department of Physics,
Tomsk State University\\
Lenin Ave. 36, Tomsk 634050, Russia}\\
\tt{kuzenko@phys.tsu.ru}
}
%\vspace{3mm}
\end{center}

%\vspace{1cm}
\begin{abstract}
We analyse the one-loop effective action of $N=4$ SYM theory in the
framework of the background field formalism in $N=2$ harmonic
superspace. For the case of on-shell background $N=2$ vector multiplet
we prove that the effective action is free of harmonic singularities.
When the lowest $N=1$ superspace component of the $N=2$ vector
multiplet is switched off, the effective action of $N=4$ SYM theory
is shown to coincide with that obtained by Grisaru et al
on the base of the $N=1$ background field method. We compute
the leading non-holomorphic corrections to the $N=4$ $SU(2)$ SYM
effective action.
\end{abstract}
\vfill

\end{titlepage}

\newpage
\setcounter{page}{1}
%\noindent

\section{Introduction}

Harmonic superspace approach \cite{gikos} is the only
manifestly $N=2$ supersymmetric formalism
developed which makes it possible to describe general $N=2$
super Yang-Mills theories in terms of unconstrained superfields
and with explicitly realized authomorphism $SU(2)_R$ symmetry
of the $N=2$ Poincar\'e superalgebra.  The Feynman rules
in harmonic superspace \cite{gios} have been successfully applied
to compute the holomorphic corrections to the effective
action of $N=2$  Maxwell multiplet coupled to the charged matter
$q$--hypermultiplet \cite{bbiko,bbik} as well as the induced hypermultiplet
self-coupling \cite{ikz}. Recently
we have developed the background field method in harmonic superspace
\cite{bbko} and applied it to rigorously prove the $N=2$
non-renormalization theorem \cite{bko}. The background field formalism
along with the results of Ref. \cite{bbiko} allowed us to derive
Seiberg's holomorphic action $\cF(\cW)$ for $N=2$ $SU(2)$
SYM theory \cite{seiberg}
for the first time
directly in $N=2$ superspace
(see Refs. \cite{dgr,n1} for $N=1$ calculations).
However, there appeared some puzzle with computing the
leading non-holomorphic correction $\cH(\cW, \bar \cW)$,
which we are going to describe a bit later.

In the Coulomb branch of $N=2$ $SU(2)$ SYM theory \cite{sw},
the effective action
$\G [\cW, \bar \cW]$ is a functional of the Abelian $N=2$ strength
$\cW$ and its conjugate $\bar \cW$. Assuming the validity of momentum
expansion, one can present $\G [\cW, \bar \cW]$
in the form
\be
\G [\cW, \bar \cW] = \left( \int {\rm d}^4 \q {\rm d}^4 \q
\cL_{eff}^{({\rm c})} + {\rm c.c.} \right) + \int {\rm d}^4 x{\rm d}^8 \q
\cL_{eff}\;.
\ee
Here the chiral effective Lagrangian $\cL_{eff}^{({\rm c})}$
is a local function of $\cW$ and its space-time derivatives,
$\cL_{eff}^{({\rm c})} = \cF(\cW) + \ldots$ ,
and the higher-derivative effective Lagrangian $\cL_{eff}$ is a
real function of $\cW$, $\bar \cW$ and their covariant derivatives,
$\cL_{eff} = \cH(\cW, \bar \cW) + \ldots$ The possibility of
non-nolomorphic quantum corrections $ \cH(\cW, \bar \cW)$
was first pointed out in \cite{hen}.
In the case of $N=4$ SYM theory
realized in $N=2$ superspace, $ \cH(\cW, \bar \cW)$ constitutes
the leading quantum correction and should have the following
structure \cite{ds} (see also \cite{dgr})
\be
\cH (\cW, {\bar \cW}) = c \,
\ln \frac{\cW^2}{\L^2} \ln \frac{ {\bar \cW}^2 }{\L^2}
\label{h1}
\ee
with some constant $c$. There are strong indications that
in $N=4$ SYM theory $\cH (\cW, {\bar \cW})$ perturbatively
is a pure one-loop effect \cite{ds} (see also \cite{bko}),
and nonperturbative corrections vanish \cite{instanton}.
The explicit value of $c$ was given in \cite{pv}
to be the result of supergraph calculations with use of the
$N=1$ background field formalism
\be
c= \frac{1}{4(4\pi)^2}\;.
\label{h2}
\ee
Indirect calculations in projective superspace
\cite{grr} led to the same value for $c$
(the authors of Ref. \cite{grr} computed in fact the low-energy
hupermultiplet action, and then identified it with that corresponding
to the vector multiplet) \footnote{The value for $c$ given
in the first version of \cite{grr} had opposite sign as compared with
(\ref{h2}).}.
What happens in harmonic superspace?

It turns out that the one-loop supergraphs contributing to
$\cH (\cW, {\bar \cW})$ in the harmonic superspace approach
contain coinciding harmonic singularities,
that is harmonic distributions at coinciding points. The problem of
coinciding harmonic singularities in the framework of
harmonic supergraph Feynman rules was first discussed in \cite{gns}.
Such singularities have no physical origin, in contrast
to ultraviolet divergences. They can appear only at intermediate stages
of calculation and should cancel each other in the final expressions
for physical quantities. But it is clear that one should very
carefully handle the relevant supergraphs in order to result in
correct amplitudes. We should stress that the problem in field
seems to be equally well characteristic of
the Feynman rules in $N=2$ projective
superspace \cite{projective}, since the structure of projective
propagators is very similar to that of harmonic ones. The origin
of this problem is an infinite number of internal degrees of freedom
associated with the bosonic internal coordinates.

To get rid of the one-loop coinciding harmonic singularities,
in \cite{bbko} we introduced,
as is generally accepted in quantum field theory,
some regularization of harmonic distributions.
Unfortunately, this regularization proves to be unsuccessful;
its use led us to the wrong conclusion  $\cH (\cW, {\bar \cW})=0$.
In a sense, the situation in hand is similar to that with
the well-known supersymmetric regularization via dimensional reduction
which leads to obstacles at higher loops. The harmonic regularization we
used turned out to be improper already at the one-loop level
\footnote{We would like to emphasize that the problem of
coinciding harmonic singularities is associated only
with perturbative calculations of the effective action and has no
direct relation to the $N=2$ background field method itself.}.

In the present paper we demonstrate that the one-loop coinciding
harmonic singularities
are easily factorized in the case with the background $N=2$ vector
multiplet on-shell. Such a choice of background superfields
is by no means critical, but it leads
to a number of technical simplifications.
When the $N=1$ chiral scalar part of the $N=2$ vector
multiplet is switched off, we show that the effective action of $N=4$ SYM,
which is derived in the framework of the background field formalism
in $N=2$ harmonic superspace, coincides with that obtained
on the base of $N=1$ background field formalism \cite{grs}. We also compute
the leading non-holomorphic corrections (\ref{h1}) to the effective action
of $N=4$ $SU(2)$ SYM theory and obtain the correct value of $c$ (\ref{h2}).

The paper is organized as follows. In section 2 we describe
the factorization of harmonic singularities and obtain a useful
representation
for the effective action of $N=4$ SYM in terms of constrained $N=2$
superfields. In section 3 we fulfil the reduction to $N=1$ superfields
and establish the fact that the two background field formalisms, in $N=2$
harmonic superspace and in $N=1$ superspace, lead to the same
one-loop effective action for $N=4$ SYM. In section 4 we compute
$\cH (\cW, {\bar \cW})$. This paper is as a direct
continuation of our work \cite{bbko}.
The conventions we follow are those of Refs. \cite{bbko,bko} with
the only exclusion: here we denote the analytic subspace measure
by ${\rm d} \z^{(-4)}$,  instead of notation ${\rm d} \z^{(-4)} {\rm d}u$
used in \cite{bbko,bko}.

\section{Elimination of harmonic singularities}

According to the results of Ref. \cite{bbko}, the purely Yang-Mills part
$\G^{(1)}[V^{++}]$
of the one-loop effective action $\G^{(1)}$ in a general $N=2$ SYM
theory is given by
\bea
\G^{(1)}[V^{++}] &=&
\frac{{\rm i}}{2}\,{\rm Tr}\,{}_{(2,2)} \, \ln
\mbox{$ {\stackrel{\frown}{\Box}}{}_\l  $}
- \frac{{\rm i}}{2}\,{\rm Tr}\,{}_{(4,0)} \, \ln
{\stackrel{\frown}{\Box}}{}_\l \non \\
& & + {\rm i}\,{\rm Tr}\,{}_{R_q}\, \ln (\nabla^{++}) +
\frac{{\rm i}}{2}\,{\rm Tr}\,{}_{R_\o}\, \ln (\nabla^{++})^2
-\frac{{\rm i}}{2}\,{\rm Tr}\,{}_{ad}\, \ln
(\nabla^{++})^2 \;.
\label{1}
\eea
Here ${\stackrel{\frown}{\Box}}$ is the analytic d'Alembertian
which reads in the $\t$-frame as follows
\bea
{\stackrel{\frown}{\Box}}{}&=&
{\cal D}^m{\cal D}_m+
\frac{{\rm i}}{2}({\cal D}^{+\alpha}W){\cal D}^-_\alpha+\frac{{\rm i}}{2}
({\bar{\cal D}}^+_{\dot\alpha}{\bar W}){\bar{\cal D}}^{-{\dot\alpha}}-
\frac{{\rm i}}{4}({\cal D}^{+\a} {\cal D}^+_\a W) \cD^{--}\non \\
&{}& +\frac{{\rm i}}{8}[{\cal D}^{+\alpha},{\cal D}^-_\alpha] W
+ \frac{1}{2}\{{\bar W},W \}\;.
\label{2}
\eea
The contributions in the first line of eq. (\ref{1}) come from the
following path integrals
\bea
\left( {\rm Det}_{(2,2)} \,
{\stackrel{\frown}{\Box}}{}_\l \right)^{-1} &=&
\int {\cal D}v^{++}  \cD u^{++}\,
\exp \left\{ -{\rm i} \; {\rm tr} \int  {\rm d}\zeta^{(-4)} \,
v^{++} {\stackrel{\frown}{\Box}}{}_\l \,u^{++} \right\} \label{3} \\
\left( {\rm Det}_{(4,0)} \,
{\stackrel{\frown}{\Box}}{}_\l \right)^{-1} &=&
\int  {\cal D}\r^{(+4)} \cD \s
\exp \left\{ -{\rm i} \; {\rm tr} \int {\rm d}\zeta^{(-4)} \,
\r^{(+4)} {\stackrel{\frown}{\Box}}{}_\l \,\s \right\}
\label{3a}
\eea
over unconstrained bosonic analytic real superfields $v^{++},\; u^{++}$
and $\r^{(+4)}, \; \s$.
The contributions in the second line  of eq. (\ref{1}) correspond
to the matter $q$--hypermultiplets (in a complex representation
$R_q$ of the gauge group) and $\o$--hupermultiplets
(in a real representation $R_{\o}$) as well as to the ghost superfields
transforming in the adjoint representation.

In the case of $N=4$ SYM theory realized in $N=2$ harmonic superspace
\cite{gios} the matter sector is formed by a single
$\o$--hypermultiplet in the adjoint representation, hence the
one-loop effective action (\ref{1}) turns into
\be
\G^{(1)}_{N=4} =
\frac{{\rm i}}{2}\,{\rm Tr}\,{}_{(2,2)} \, \ln
{\stackrel{\frown}{\Box}}{}_\l
- \frac{{\rm i}}{2}\,{\rm Tr}\,{}_{(4,0)} \, \ln
{\stackrel{\frown}{\Box}}{}_\l \;.
\label{4}
\ee
A simple inspection of harmonic supergraphs shows that both contributions
to $\G^{(1)}_{N=4}$ contain harmonic singularities.
On general grounds, the singular parts
of ${\rm Tr}\,{}_{(2,2)} \, \ln {\stackrel{\frown}{\Box}}{}_\l $
and ${\rm Tr}\,{}_{(4,0)} \, \ln {\stackrel{\frown}{\Box}}{}_\l $
should cancel each other. Complete factorization of harmonic singularities
can be most simply established in the
case when the background $N=2$ gauge superfield is chosen on-shell
\be
\cD^{\a (i} \cD^{j)}_\a  W = 0
\label{5}
\ee
what we assume below. A more complete analysis will be given somewhere
else.
Under requirement (\ref{5}) the operators $\cD^{++}$ and
${\stackrel{\frown}{\Box}}$, which move every
analytic superfield into analytic ones, commute as a consequence
of the identity (given in the $\t$-frame)
\be
[\cD^{++}, {\stackrel{\frown}{\Box}}] \F^{(q)} =
\frac{{\rm i}}{4} \,(1-q)(\cD^{+ \a} \cD^+_\a W) \,  \F^{(q)}
\ee
for an arbitrary analytic superfield $\F^{(q)}$ with $U(1)$ charge $q$.

Let us consider the following non-degenerate replacement of variables
\bea
v^{++} &=& \cF^{++} + \nabla^{++} \s \non \\
u^{++} &=& \cG^{++} + \nabla^{++} \int {\rm d} \tilde{\z}^{(-4)}\;
{\bf G}^{(0,0)}(\z, \tilde{\z}) \r^{(+4)}(\tilde{\z})
\label{6}
\eea
where $v^{++},\; u^{++}$ and $\s,\; \r^{(+4)}$ are unconstrained
analytic real superfields, while the analytic real superfields
$\cF^{++}$ and $\cG^{++}$ are constrained to be covariantly linear
\be
\nabla^{++} \cF^{++} = 0 \qquad \nabla^{++} \cG^{++} = 0\;.
\label{7}
\ee
The Green's function ${\bf G}^{(0,0)}(\z_1, \z_2)$ is the Feynman
propagator of $\o$--hypermultiplet coupled to the $N=2$ gauge
superfield. It satisfies the equation
\be
(\nabla_1^{++})^2 {\bf G}^{(0,0)}(1, 2) = - \d_A^{(4,0)}(1,2)
\ee
and reads most simply in the $\t$--frame \cite{bko}
\be
{\bf G}^{(0,0)}_\t(1, 2)
=  \frac{1}{{\stackrel{\frown}{\Box}}{}_1}
{\stackrel{\longrightarrow}{(\cD_1^+)^4}}{}
\left\{ \delta^{12}(z_1-z_2)
{(u^-_1 u^-_2)\over (u^+_1 u^+_2)^3}
\right\}
{\stackrel{\longleftarrow}{(\cD_2^+)^4}} \;.
\label{8}
\ee

We fulfil the replacement of variables (\ref{6}) in
the path integral
\be
1 = \int {\cal D}v^{++}  \cD u^{++}\,
\exp \left\{ {\rm i} \; {\rm tr} \int  {\rm d}\zeta^{(-4)} \,
v^{++} \,u^{++} \right\}\;.
\label{9}
\ee
Denoting by $J$ the corresponding Jacobian, we get
\bea
1 &=& J\;
\int  {\cal D}\r^{(+4)} \cD \s
\exp \left\{ {\rm i} \;  {\rm tr}  \int {\rm d} \zeta^{(-4)} \,
\r^{(+4)}  \,\s \right\} \non \\
& & \times
\int  {\cal D}\cF^{++} \cD \cG^{++}
\exp \left\{ {\rm i} \;{\rm tr} \int {\rm d}\zeta^{(-4)} \,
\cF^{++}  \,\cG^{++} \right\} \non \\
&=& J   \;
\int  {\cal D}\cF^{++} \cD \cG^{++}
\exp \left\{ {\rm i} \; {\rm tr} \int {\rm d}\zeta^{(-4)} \,
\cF^{++}  \,\cG^{++} \right\} \;.
\label{10}
\eea
Now, we fulfil the same replacement of variables in (\ref{3})
\bea
\left( {\rm Det}_{(2,2)} \,
{\stackrel{\frown}{\Box}}{}_\l \right)^{-1} &=& J\;
\int  {\cal D}\r^{(+4)} \cD \s
\exp \left\{ -{\rm i} \;{\rm tr} \int {\rm d}\zeta^{(-4)} \,
\r^{(+4)} {\stackrel{\frown}{\Box}}{}_\l \,\s \right\}\non \\
& & \times
\int  {\cal D}\cF^{++} \cD \cG^{++}
\exp \left\{ -{\rm i} \; {\rm tr} \int {\rm d}\zeta^{(-4)} \,
\cF^{++} {\stackrel{\frown}{\Box}}{}_\l \,\cG^{++} \right\} \;.
\label{11}
\eea
Expressing here $J$ as in eq. (\ref{10}) and making use of (\ref{3a}),
we then result in
\be
\exp \{ 2 {\rm i} \, \G^{(1)}_{N=4} \}=
\frac{
\int  {\cal D}\cF^{++} \cD \cG^{++}
\exp \left\{ -{\rm i} \; {\rm tr} \int {\rm d}\zeta^{(-4)} \,
\cF^{++} {\stackrel{\frown}{\Box}}{}_\l \,\cG^{++} \right\} }
{
\int  {\cal D}\cF^{++} \cD \cG^{++}
\exp \left\{ {\rm i} \;  {\rm tr} \int {\rm d}\zeta^{(-4)} \,
\cF^{++}  \,\cG^{++} \right\} } \;.
\label{12}
\ee

It is not difficult to obtain another representation for
the effective action
\be
\exp \{  {\rm i} \, \G^{(1)}_{N=4} \}=
\frac{
\int  {\cal D}\cF^{++}
\exp \left\{- \frac{{\rm i}}{2} \; {\rm tr} \int {\rm d}\zeta^{(-4)} \,
\cF^{++} {\stackrel{\frown}{\Box}}{}_\l \,\cF^{++} \right\} }
{
\int  {\cal D}\cF^{++}
\exp \left\{ \frac{{\rm i}}{2} \;{\rm tr} \int {\rm d}\zeta^{(-4)} \,
\cF^{++}  \,\cF^{++} \right\} }
\label{13}
\ee
which will be the starting point for our further analysis.

Both functional integrals in the r.h.s. of (\ref{13}) are
carried out over constrained analytic superfields depending on
the background gauge superfield. In particular, the functional
measure $\cD \cF^{++}$ depends nontrivially on the gauge superfield.
But the relevant constraint
\be
\nabla^{++} \cF^{++} = D^{++} \cF^{++} + {\rm i}\,
[V^{++}, \cF^{++}] =0
\label{14}
\ee
can be solved in terms of a $V^{++}$ independent linear analytic
superfield $F^{++}$
\be
D^{++} F^{++} = 0
\label{anal}
\ee
as a power series in $V^{++}$
\be
\cF^{++} = F^{++} + \sum_{n=1}^{\infty} A^n F^{++}
\label{15}
\ee
where the operator $A$ transforms any analytic superfield $\f^{++}(\z)$
into an analytic one and is defined by
\be
A \, \f^{++} (\z) = {\rm i}\; D^{++}
\int {\rm d} \tilde{\z}^{(-4)} {\rm G}^{(0,0)}(\z, \tilde{\z})
[ V^{++}(\tilde{\z}), \f^{++}(\tilde{\z})]\;.
\label{16}
\ee
Here ${\rm G}^{(0,0)}(1,2)$ is the free $\o$--hypermultiplet
propagator \cite{gios} and it can be obtained from (\ref{8}) by switching off
the background
gauge superfield.
Of course, the functional integration measures $\cD \cF^{++}$ and
$\cD F^{++}$ are related by some $V^{++}$ dependent Jacobian, but the
latter drops out of the r.h.s. of (\ref{13}). Therefore, we can
simply replace $\cD \cF^{++}$ in the r.h.s. of (\ref{13}) by $\cD F^{++}$
and then pertubatively compute both functional integrals by
properly defining the relevant free propagators. This is a simple
task after heroic efforts undertaken
by the projective superspace group \cite{projective} on working out
the propagators
of constrained superfields. For example, the propagator
corresponding to a free tensor multiplet with action \cite{gio}
\be
S = \hf \int {\rm d} \z^{(-4)} (F^{++})^2
\ee
reads
\bea
&&<F^{++}(1)\, F^{++}(2)> = {\rm i}\; \Pi^{(2,2)} (1,2) \non \\
&&\Pi^{(2,2)} (1,2) = - \frac{ (D^+_1)^4 (D^+_2)^4 }{\Box_1}
\frac{1}{(u^+_1 u^+_2)^2} \d^{12} (z_1 - z_2)
\label{17}
\eea
with $\Pi^{(2,2)} (1,2)$ the projector operator \cite{gios} for the
analytic superfields constrained by (\ref{anal}).

The main advantage of representation (\ref{13}) is that it contains
no harmonic singularities. The integration in (\ref{13}) is in fact
carried out over ordinary $N=2$ constrained superfields coupled
to the super Yang-Mills multiplet. This becomes obvious in the
$\t$--frame in which $\cF^{++}$ reads
\be
\cF^{++}_\t = \cF^{ij} (z) u^+_i u^+_j
\label{18}
\ee
where the $u$--independent isovector $\cF^{ij}$ is real
\be
\overline{\cF^{ij} } = \cF_{ij}
\label{19}
\ee
and satisfies the constraints
\be
\cD_\a^{(i} \cF^{jk)} = {\bar \cD}_\ad^{(i} \cF^{jk)} =0 \;.
\label{20}
\ee
Because of (\ref{5}), the operator ${\stackrel{\frown}{\Box}}$
moves the space of such superfields into itself
\bea
{\stackrel{\frown}{\Box}} {} \cF^{ij} &=&
\left( \cD^a \cD_a + \hf \{W, \bar W \} \right) \cF^{ij} \non \\
&& + \frac{{\rm i}}{3}(\cD^{\a (i} W)\, {\bf \mbox{$\cdot $} } \,
\cD_{\a |k|} \cF^{j) k} +  \frac{{\rm i}}{3}
{\bar \cD}_\ad^{(i} {\bar W} \, {\bf \mbox{$\cdot $} } \,
{\bar \cD}^\ad_{|k|} \cF^{j) k} \;.
\label{21}
\eea
At the same time the representation (\ref{13}) allows us to make a
simple reduction to $N=1$ superspace.

\section{Reduction to N = 1 superfields}

We introduce the Grassmann coordinates of $N=1$ superspace
$(\q^\a, {\bar \q}_\ad)$ as part
of those $(\q^\a_i, {\bar \q}^j_\ad)$ parametrizing its $N=2$ extension
\be
\q^\a = \q^\a_1 \qquad {\bar \q}_\ad = {\bar \q}^1_\ad
\non
\ee
and define $N=1$ projections of $N=2$ superfields by the standard
rule
\be
U| = U(x^m, \q^\a_i, {\bar \q}^j_\ad)|_{\q_2={\bar \q}^2 =0}\;.
\non
\ee
The $N=1$ gauge covariant derivatives are
\be
\cD_\a = \cD^1_\a | = D^1_\a + {\rm i}\; \cA^1_\a | \qquad
{\bar \cD}^\ad =  {\bar \cD}^\ad_1| = {\bar D}^\ad_1 +
{\rm i}\; {\bar \cA}_1^\ad |\;.
\non
\ee
The covariantly chiral $N=2$ strength $W$ leads to the two
$N=1$ superfields
\bea
\F = W|  & \qquad &     {\bar \cD}_\ad \F =0 \non \\
2{\rm i} \; W_\a = \cD^2_\a W| & \qquad  &
{\bar \cD}_\ad W_\a =0
\label{22}
\eea
which are covariantly chiral, and $W_\a$ satisfies the Bianchi identity
\be
\cD^\a W_\a =  {\bar \cD}_\ad {\bar W}^\ad \;.
\label{23}
\ee
The algebra of $N=1$ derivatives reads
\bea
&& \{ \cD_\a , {\bar \cD}_\ad \} = -2 {\rm i}\; \cD_{\a \ad} \qquad
\;\;\;
\{ \cD_\a ,  \cD_\b \} = \{ {\bar \cD}_\ad , {\bar \cD}_\bd \} =0 \non \\
&&[ \cD_{\a \ad} , \cD_\b ] = -2 {\rm i}\; \varepsilon_{\a \b}
{\bar W}_\ad \qquad
[ \cD_{\a \ad} , {\bar \cD}_\bd ] = -2 {\rm i} \;
\varepsilon_{\ad \bd} W_\a \;.
\label{24}
\eea
One of the $N=1$ counterparts of requirement (\ref{5}) is
\be
\cD^\a W_\a = [{\bar \F}, \F] \;.
\label{25}
\ee

The $N=1$ projections of the superfield
$\cF^{ij}$ constrained by eqs. (\ref{19}) and (\ref{20})
are
\be
\J = \cF^{22}| \qquad {\bar \J} = \cF^{11}| \qquad
F = {\bar F}= -2{\rm i} \cF^{12}|
\label{26}
\ee
and satisfy the constraints
\be
{\bar \cD}_\ad \J = 0 \qquad
-\frac{1}{4} {\bar \cD}^2 F + [ \F , \J ] =0\;.
\label{27}
\ee
Therefore, $\J$ is a covariantly chiral superfield, while the real
superfield $F$ is subject to a modified linear constraint.

To reduce the actions arising in the r.h.s. of (\ref{13}) to $N=1$
superfields, we can go on as follows. Let $L^{(+4)}(\z)$ be a
gauge invariant real
analytic superfield constrained by
\be
D^{++}  L^{(+4)} = 0\;.
\ee
Such a superfield can be represented in the form
\be
L^{(+4)}(\z) = L^{ijkl}(z) u^+_i u^+_j u^+_k u^+_l
\ee
where the $u$--independent superfield $ L^{ijkl}$ satisfies the
constraints
\be
D_\a^{(i_1} L^{i_2 \cdots i_5)} =
{\bar D}_\ad^{(i_1} L^{i_2 \cdots i_5)} =0\;.
\ee
These constraints imply
\be
\int {\rm d} \z^{(-4)} L^{(+4)} = 6 \int {\rm d}^8 z L^{1122}| \qquad
{\rm d}^8 z = {\rm d}^4 x {\rm d}^2 \q {\rm d}^2 {\bar \q}
\label{action}
\ee
with ${\rm d}^8 z$ being the full $N=1$ superspace measure. In the role
of $L^{(+4)}$ we choose
\be
L^{(+4)} = {\rm tr}\, (\cF^{++} \cG^{++}) =
{\rm tr}\, (\cF^{++}_\t \cG^{++}_\t)
\ee
with analytic superfields $\cF^{++}$ and $\cG^{++}$ constrained as in
(\ref{7}). Then we get
\be
{\rm tr} \int {\rm d} \z^{(-4)}
\cF^{++} \cG^{++}
 = {\rm tr} \int {\rm d}^8 z \left( \cF^{11}| \, \cG^{22}|
+ \cG^{11}|\, \cF^{22}| + 4 \cF^{12}|\,\cG^{12}| \right)\;.
\ee

To compare our $N=2$ superspace results for the one-loop effective
action of $N=4$ super Yang-Mills theory with those derived in $N=1$
superspace \cite{grs}, we switch off the
lowest $N=1$ superspace component of the $N=2$ vector multiplet  by setting
\be
\F=0\;.
\label{off}
\ee
Then eq. (\ref{21}) takes the form
\be
\left({\stackrel{\frown}{\Box}} {} \cF^{ij} \right)|
= {\stackrel{\sim}{\Box}} {}  \cF^{ij}| \qquad \quad
 {\stackrel{\sim}{\Box}} {}
= \cD^a \cD_a  - W^\a \cD_\a  + {\bar W}_\ad {\bar \cD}^\ad\;.
\label{box}
\ee
As is seen, the operator
${\stackrel{\sim}{\Box}} {} $ does not mix the components of
$\cF^{ij}|$. It is not surprising, but nevertheless very important that
${\stackrel{\sim}{\Box}} {} $ is exactly the operator which enters
the quadratic part of quantum gauge superfield action in
the $N=1$ background field approach \cite{grs}.
Now, eq. (\ref{13}) turns into
\be
\exp \{  {\rm i} \, \G^{(1)}_{N=4} \}=
\frac{
\int  \cD {\bar \J} \cD  \J \cD F
\exp \left\{ {\rm i} \; {\rm tr} \int {\rm d}^8 z \, \left(
- {\bar \J} {\stackrel{\sim}{\Box}} {}  \J
+ \hf F {\stackrel{\sim}{\Box}} {}  F \right) \right\}
}
{
\int  \cD {\bar \J} \cD  \J \cD F
\exp \left\{ {\rm i} \;{\rm tr} \int {\rm d}^8 z \, \left(
{\bar \J} \J - \hf F^2 \right) \right\} } \;.
\label{28}
\ee
It should be remarked that here $F$ is a covariantly linear superfield,
as a consequence of eqs. (\ref{27}) and (\ref{off}).

Let us recall the path-integral representation for $\G^{(1)}_{N=4}$
derived in the framework of $N=1$ background field method \cite{grs}
\be
\exp \{  {\rm i} \, \G^{(1)}_{N=4} \}=
\int  \cD U  \exp \left\{
\frac{{\rm i}}{2} \;{\rm tr} \int {\rm d}^8 z \, U
{\stackrel{\sim}{\Box}} {}  U \right\}
\label{n1}
\ee
where the integration variable $U$ is an unconstrained real superfield.
It is not difficult to see that the representations (\ref{28}) and
(\ref{n1}) are equivalent. Really, because of eqs. (\ref{25}) and
(\ref{off}) we have a
well defined and gauge covariant decomposition of $U$ into a sum of
chiral--antichiral and linear parts
\be
U = \J + {\bar \J} + F \qquad {\bar \cD}_\ad \J =0 \qquad
{\bar \cD}^2 F = 0\;.
\label{rv}
\ee
Jacobian $J$ of such a replacement of variables can be read off from
\bea
1 &=& \int  \cD U  \exp \left\{
\frac{{\rm i}}{2} \;{\rm tr} \int {\rm d}^8 z \, U^2 \right\} \non \\
 &=& J
\int  \cD {\bar \J} \cD  \J \cD F
\exp \left\{ {\rm i} \;{\rm tr} \int {\rm d}^8 z \, \left(
{\bar \J} \J + \hf F^2 \right) \right\}  \;.
\eea
If we fulfil the replacement of variables (\ref{rv}) in (\ref{n1})
and represent the corresponding Jacobian as in the relation just
obtained, we will end up with (\ref{28}).

Another interesting representation for $\G^{(1)}_{N=4}$ can be obtained
when $\F$ is covariantly constant and lies along a flat direction of
the $N=2$ SYM potential
\be
\cD_\a \F = 0 \qquad [{\bar \F}, \F] = 0
\ee
what is consistent only if
\be
[W_\a , \F] = [{\bar W}_\ad , \F] = 0\;.
\ee
In this case we have
\be
\left({\stackrel{\frown}{\Box}} {} \cF^{ij} \right)|
= \Delta \cF^{ij}| \qquad \quad
 \Delta = \cD^a \cD_a  - W^\a \cD_\a  + {\bar W}_\ad {\bar \cD}^\ad
+\hf \{ {\bar \F}, \F\} \;.
\label{del}
\ee
The background gauge freedom can be used to make $\F$
constant. Then we stay with unbroken gauge transformations leaving
$\F$ invariant. Let us for simplicity consider $SU(2)$ in the role
of gauge group (with the generators $\t^{\underline{a}}  =
\frac{1}{\sqrt{2}} \s^{\underline{a}}$) and choose the $N=2$ strength
$W$  in the $z$-direction
\be
W = W^{\underline{3}} \t^{\underline{3}} \equiv \cW \t^{\underline{3}}
\qquad \F \equiv \f \t^{\underline{3}} \qquad
W_\a \equiv \cW_\a \t^{\underline{3}} \;.
\ee
Since only the $U(1)$ subgroup generated by $\t^{\underline{3}}$
is gauged, the $z$-components of all quantum superfields
$F^{\underline{a}}$, $\J^{\underline{a}}$ and
${\bar \J}^{\underline{a}}$ do not interact with the background
superfields and, hence, completely decouple. It is useful to combine
the rest components into ones having definite charge with respect to
$\t^{\underline{3}}$
\bea
 V= F^{\underline{1}} - {\rm i} F^{\underline{2}} &\qquad&
\t^{\underline{3}} V = e V\non \\
\J_\pm = \J^{\underline{1}} \mp {\rm i} \J^{\underline{2}} &\qquad&
\t^{\underline{3}} \J_\pm = \pm e \J_\pm
\eea
with $e = \sqrt{2}$. Now, eq. (\ref{27}) tells us
\be
\frac{1}{4} {\bar \cD}^2 V = \f e \J_+ \qquad
\frac{1}{4} {\bar \cD}^2 {\bar V} = - \f e \J_- \;.
\ee
These relations can be treated as the definition of chiral superfields
$\J_+$ and $\J_-$ in terms of an unconstrained complex scalar
superfield $V$ and its conjugate. As a result, eq. (\ref{13}) becomes
%(after some rescaling of the quantum superfields)
\be
\exp \{  {\rm i} \, \G^{(1)}_{N=4} \}=
\frac{
\int  \cD {\bar V}  \cD V
\exp \left\{ -\frac{{\rm i}}{2} \; {\rm tr} \int {\rm d}^8 z \,
{\bar V} \bB \Delta V \right\}
}
{
\int  \cD {\bar V} \cD V
\exp \left\{ \frac{{\rm i}}{2} \;{\rm tr} \int {\rm d}^8 z \,
{\bar V} \bB V \right\} }
\label{40}
\ee
where $\bB$ denotes the following non-singular operator
\be
\bB = \frac{1}{16} \{ {\bar \cD}^2, \cD^2 \} - e^2 |\f|^2 \;.
\ee
Since $V$ is unconstrained, from here we immediately get
\be
\exp \{  {\rm i} \, \G^{(1)}_{N=4} \}=
\int  \cD {\bar V}  \cD V
\exp \left\{ \frac{{\rm i}}{2} \; {\rm tr} \int {\rm d}^8 z \,
{\bar V} \Delta V \right\}
\label{final}
\ee
where $\Delta$ now reads
\be
\Delta = \cD^a \cD_a - e \cW^\a \cD_\a + e {\bar \cW}_\ad {\bar \cD}^\ad
+ e^2 |\f|^2 \;.
\label{del2}
\ee
The structure of operator $\Delta$ is such that
$\G^{(1)}_{N=4}$ can be computed in the framework of the $N=1$ superfield
Schwinger-DeWitt technique \cite{bk}.

\section{Non-holomorphic corrections in N = 4 SYM}

In the present we compute the leading non-holomorphic contribution
to the one-loop effective action of $N=4$ $SU(2)$ super Yang-Mills
theory
\be
\G = \int {\rm d}^{12} z \, \cH (\cW, {\bar \cW})
\label{nh}
\ee
where the Abelian $N=2$ strength $\cW$ corresponds to the unbroken
$U(1)$ subgroup of $SU(2)$ in the Coulomb branch. $\G$ can be readily
reduced to $N=1$ superfields. Introducing the $N=1$ projections
of $\cW$
\be
\f = \cW| \qquad 2{\rm i} \,\cW_\a = \cD_\a^2 \cW|
\ee
one gets (see Refs. \cite{dgr,hen} for more detail)
\be
\G =  \int {\rm d}^{8} z \, \cW^\a \cW_\a {\bar \cW}_\ad  {\bar \cW}^\ad \;
\frac{\pa^4 \cH (\f, {\bar \f})}{\pa \f^2 \pa {\bar \f}^2} \;+\;
\cdots
\label{nh1}
\ee
where the dots mean the terms of third and lower orders in $\cW_\a$
and $\bar \cW_\ad$.
To compute the structure present here, it is sufficient to
evaluate $\G^{(1)}_{N=4}$ for constant $\f$ and $\cW_\a$,
and such a background is consistent with our restriction (\ref{5})
used throughout the present paper.

In accordance with eqs. (\ref{final}) and (\ref{del2}), we have
\be
\G^{(1)}_{N=4} = -{\rm i} \int_{0}^{\infty} \frac{{\rm d} s}{s}
{\rm e}^{-{\rm i}(e^2 |\f|^2 - {\rm i} \varepsilon )s}
\int {\rm d}^8 z \,\cU (z,z|s)
\ee
where $\cU (z,z'|s)$ denotes the Schwinger's kernel \cite{bk}
\be
\cU (z,z'|s) = \exp \left\{ - {\rm i} s \left(
\cD^a \cD_a - e \cW^\a \cD_\a + e {\bar \cW}_\ad {\bar \cD}^\ad \right)
\right\} \d^8 (z-z') \;.
\ee
For computing $\G$ given by eq. (\ref{nh1}) it is sufficient to use
the approximation
\be
\cU (z,z'|s)
 \approx   \exp \left\{ {\rm i} s \left( e \cW^\a D_\a -
e {\bar \cW}_\ad {\bar D}^\ad \right) \right\} \cU_0 (z,z'|s)
\ee
with $\cU_0 (z,z'|s) $ the free Schwinger's kernel \cite{bk}
\be
\cU_0 (z,z'|s)= {\rm e}^{-{\rm i} s \pa^a \pa_a} \d^8 (z-z')
= \frac{{\rm i}}{(4\pi {\rm i}s)^2} \d^4(\q - \q')
{\rm e}^{-{\rm i} (x-x')^2/4s}\;.
\ee
Using the identity
$$
\frac{1}{16}D^2 {\bar D}^2 \d^4 (\q-\q') = 1
$$
we obtain
\bea
\G^{(1)}_{N=4} &\approx& \frac{e^4}{(4\pi)^2}
\int {\rm d}^8 z \;\cW^\a \cW_\a {\bar \cW}_\ad  {\bar \cW}^\ad
\int_{0}^{\infty} {\rm d} s\,s\;
{\rm e}^{-s\, e^2 |\f|^2 } \non \\
&=& \frac{1}{(4\pi)^2}
\int {\rm d}^8 z\;
\cW^\a \cW_\a {\bar \cW}_\ad  {\bar \cW}^\ad
\frac{1}{\f^2 {\bar \f}^2 }
\eea
where all terms involving derivatives of $\cW_\a$ and ${\bar \cW}_\ad$
have been omitted. Comparing the relation obtained with (\ref{nh1}),
we get
\be
\frac{\pa^4 \cH (\cW, {\bar \cW}) }{\pa \cW^2 \pa {\bar \cW}^2} =
\frac{1}{(4\pi)^2} \frac{1}{\cW^2 {\bar \cW}^2 } \;.
\ee
A general solution of this equation (modulo purely holomorphic
terms) reads
\bea
\cH (\cW, {\bar \cW})  &=&  \frac{1}{(4\pi)^2} \ln  \frac{\cW}{\L}
 \ln \frac{{\bar \cW}}{\L} \non \\
&+&   \cW {\bar f}({\bar \cW})  +  f(\cW ) {\bar \cW}
\eea
with some scale $\L$ and some holomorphic function $f(\cW)$.
The structures in the second line
are in conflict with the scale and chiral invariances of $N=4$ SYM theory.
Moreover, all contributions to $\cH (\cW, {\bar \cW})$ coming from
the $N=2$ supergraphs involve equal powers of $\cW$ and $\bar \cW$.
%(this does not, however, rule out the quadratic combination
%$\cW {\bar \cW}$).
Therefore, we conclude
\be
\cH (\cW, {\bar \cW}) =  \frac{1}{(4\pi)^2 } \, \ln  \frac{\cW}{\L}
\ln \frac{{\bar \cW}}{\L}
= \frac{1}{4(4\pi)^2 } \,
\ln \frac{\cW^2}{\L^2} \ln \frac{ {\bar \cW}^2 }{\L^2} \;.
\ee
Our final expression for $\G$ coincides with the result of $N=1$
calculations announced in \cite{pv}  as well as  with that
obtained by indirect methods in projective superspace \cite{grr}.
The overall sign for $\cH (\cW, {\bar \cW})$
can be readily determined, if we compare the coefficient for
$(\cW {\bar \cW})^2$
in a power series expansion of $\cH (\cW, {\bar \cW})$,
with respect to some point $\cW_0$, with similar $N=1$ quantum corrections
quadric in $\cW_\a$ and ${\bar \cW}_\ad$ which were found in
$Superspace$ \cite{ggrs}.
In a separate paper we are going to compute $\cH (\cW, {\bar \cW})$
directly in $N=2$ harmonic superspace and without any restrictions
on the background $N=2$ gauge superfield.

\vspace{1cm}

\noindent
{\bf Acknowledgements.}
The authors are grateful to E.I. Buchbinder,
M.T. Grisaru, E.A. Ivanov, B.A. Ovrut and
E. Sokatchev for valuable
discussions. We acknowledge a partial support from RFBR grant,
project No 96-02-1607; RFBR-DFG grant, project No 96-02-00180;
INTAS grant, INTAS-96-0308; grant in the field of fundamental
natural sciences from Ministry of General and Professional
Education of Russian Federation.

\end{document}